\documentclass{article}

\usepackage{microtype}
\usepackage{graphicx}
\usepackage{subfigure}
\usepackage{booktabs} 

\usepackage{hyperref}
\usepackage{amsmath}
\usepackage{algorithmic}

\usepackage{amsmath,amsfonts,bm,bbm}

\def\eqref#1{equation~(\ref{#1})}

\def\1{\bm{1}}

\usepackage{cancel}

\usepackage{listings}
\usepackage{enumerate}

\DeclareMathAlphabet{\mathsfit}{\encodingdefault}{\sfdefault}{m}{sl}
\SetMathAlphabet{\mathsfit}{bold}{\encodingdefault}{\sfdefault}{bx}{n}

\usepackage{amsthm}

\makeatletter
\newtheorem*{rep@theorem}{\rep@title}
\newcommand{\newreptheorem}[2]{%
\newenvironment{rep#1}[1]{%
 \def\rep@title{#2 \ref{##1}}%
 \begin{rep@theorem}}%
 {\end{rep@theorem}}}
\makeatother

\newreptheorem{lemma}{Lemma}

\usepackage[accepted]{icml2021}

\begin{document}

\twocolumn[
\icmltitle{A Non-Negative Matrix Factorization Game}



\icmlsetsymbol{equal}{*}

\begin{icmlauthorlist}
\icmlauthor{Satpreet~H. Singh}{uw}
\end{icmlauthorlist}


\icmlaffiliation{uw}{University of Washington, Seattle, WA, USA, 98195}
\icmlcorrespondingauthor{Satpreet~H. Singh}{satsingh@uw.edu}

\icmlkeywords{Game Theory, Matrix Factorization, Machine Learning, Data Science}

\vskip 0.3in
]



\printAffiliationsAndNotice{}  

\begin{abstract}
    We present a novel game-theoretic formulation of Non-Negative Matrix Factorization (NNMF), a popular data-analysis method with many scientific and engineering applications. 
    The game-theoretic formulation is shown to have favorable scaling and parallelization properties, while retaining reconstruction and convergence performance comparable to the traditional Multiplicative Updates \cite{lee1999learning} algorithm.
\end{abstract}

\section{Introduction}
Non-Negative Matrix Factorization (NNMF) is a matrix decomposition method that approximates a low-rank non-negative matrix $\mathbf{X} \in \mathbb{R}_{\geq 0}^{I \times J}$, using two non-negative matrices 
$\mathbf{W} \in \mathbb{R}_{\geq 0}^{I \times K},$ and 
$\mathbf{H} \in \mathbb{R}_{\geq 0}^{K \times J},$ of predetermined rank $K$ according to:

\begin{align}
\mathbf{X} \simeq \mathbf{W} \mathbf{H}
\end{align}

The non-negative rank $K$ of the non-negative matrix $\mathbf{X}$ can be higher than the traditional matrix rank over the real field. 

NNMF has a long history of use in both scientific and engineering settings due to it's tendency to reveal interesting properties about the underlying data. 
For example, \citet{lee1999learning} popularized NNMF after applying it to natural images, showing that it can learn meaningful visual representations "the parts of objects”. 
NNMF has also been applied to large-scale textual corpora to learn topic-models \cite{lee1999learning, pauca2004text} and to high-throughput omics data for time-course analysis \cite{stein2018enter}. 

Recently \citet{gemp2020eigengame, gemp2021eigengame}, reformulated a data-analysis algorithm called Principal Components Analysis (PCA) \cite{jolliffe2002principal} as a $K$-player game, to show that it can be be massively scaled and implemented in a distributed manner.
Inspired by their work, we present a novel game-theoretic formulation of Non-Negative Matrix Factorization (NNMF) and provide an empirical analysis of our formulation against a traditional NNMF algorithm \cite{lee1999learning}.

\section{PCA through Nash Equilibrium}
Principal Components Analysis (PCA) is a data-analysis algorithm that uses an orthogonal linear transformation to transform a set of observations $\mathbf{X} \in \mathbb{R}^{n \times d}$ to a new coordinate system, such that the projections of the data onto the new coordinates have decreasing variance.
In other words, the first coordinate (known as the first \textit{principal component}) is the direction of maximal variance of the data, the second (orthogonal) coordinate is the direction of second-highest variance of the data, and so on. 
[See \citet{jolliffe2002principal} for a comprehensive review.]

PCAs are typically calculated in one of two ways:
(1) using the eigen-decomposition of the covariance matrix $\Sigma$ of the data matrix $ \mathbf{X}$ or 
(2) using the singular value decomposition (SVD) of the data matrix $ \mathbf{X}$.

Using the former method, let the data matrix be $\mathbf{X} \in \mathbb{R}^{n \times d}$, and it's covariance matrix be $M ^{d \times d} = \frac{1}{n} \mathbf{X}^{T} \mathbf{X}$.
PCA then reduces to the eigenvalue problem $M V = V \Lambda$, where $\Lambda$ is a diagonal matrix comprised of eigenvalues (known as PC \textit{loadings}) and the columns of the orthonormal matrix $V$ provides the sought PC directions.

\subsection{Game-theoretic PCA formulation}

In a recent work, \citet{gemp2020eigengame} showed how the top-$k$ right singular vectors of an $n$-observation data matrix $\mathbf{X} \in \mathbb{R}^{n \times d}$ can be calculated by a $k$-player game. 
Alternatively, if one is working with a mini-batch $\mathbf{X_t}$ of $n' < n$ observations, the top-$k$ right singular vectors are given by the top-$k$ eigenvectors of the positive semidefinite sample covariance matrix $\Sigma = \mathbb{E}[\,\frac{1}{n'} \mathbf{X_t}^{T} \mathbf{X_t} ]\, $.

In their game formulation, each of $k$ players receives access to a minibatches of the data and `owns' one of the $k$ (approximate) eigenvectors $\hat{v}_i$.
Each player then tries to maximize a utility function $u_i(\hat{v}_i \vert \hat{v}_{j < i})$ that is unique to them:
\begin{align}
   \max_{\hat{v}_i^\top \hat{v}_i = 1} \Big\{ u_i(\hat{v}_i \vert \hat{v}_{j < i}) &= \overbrace{\hat{v}_i^\top \Sigma \hat{v}_i}^{\text{Variance}} - \sum_{j < i} \overbrace{\frac{\langle \hat{v}_i, \Sigma \hat{v}_j \rangle^2}{\langle \hat{v}_j, \Sigma \hat{v}_j \rangle}}^{\perp\text{-penalty}} \Big\}
\end{align}

Here, all expressions involving the inequality $(j < i)$ indicate that players with a lower index (`parents') have already played before the current (`child') player.
As shown above, utility functions consist of two terms that balance competing objectives: one that tries to maximize the variance captured by the player's eigenvector, and the other that tries to maximally orthogonalize this eigenvector to all parent eigenvectors.
To reduce underspecification, eigenvectors are restricted to lie on the unit sphere ($\hat{v}_i^{(t)} \in \mathcal{S}^{d-1}$).

\begin{algorithm}[tbhp!]
\begin{algorithmic}[1]
    \STATE Given: data minibatches $\mathbf{X_t} \in \mathbb{R}^{m \times d}$, total iterations $T$, initial vector $\hat{v}_i^0 \in \mathcal{S}^{d-1}$ (unit sphere in $d-1$ dimensions), and step size $\alpha$.
    \STATE $\hat{v}_i \leftarrow \hat{v}_i^0$
    \FOR{$t = 1: T$}
        \STATE $\nabla_{\hat{v}_i} \leftarrow 2 \mathbf{X_t}^\top \Big[ \mathbf{X_t} \hat{v}_i - \sum_{j < i} \frac{\langle \mathbf{X_t} \hat{v}_i, \mathbf{X_t} \hat{v}_j \rangle}{\langle \mathbf{X_t} \hat{v}_j, \mathbf{X_t} \hat{v}_j \rangle} \mathbf{X_t} \hat{v}_j \Big]$
        \STATE $\nabla^R_{\hat{v}_i} \leftarrow \nabla_{\hat{v}_i} - \langle \nabla_{\hat{v}_i}, \hat{v}_i \rangle \hat{v}_i$
        \STATE $\hat{v}_i' \leftarrow \hat{v}_i + \alpha \nabla^R_{\hat{v}_i}$
        \STATE $\hat{v}_i \leftarrow {\hat{v}_i'}/{|| \hat{v}_i' ||}$
        \STATE \texttt{broadcast}($\hat{v}_{i}$) to all other players
    \ENDFOR
    \STATE return $\hat{v}_i$
\end{algorithmic}
\caption{PCA game \cite{gemp2020eigengame}}
\label{algo:eigengame}
\end{algorithm}

The complete algorithm is given in Algorithm \ref{algo:eigengame}.
At each iteration, each player updates their eigenvector $\hat{v}_i$ using a fixed step size \textit{Riemannian} gradient ascent update $\nabla^R_{\hat{v}_i}$, that ensures that each iterate lies on the acceptable manifold \cite{bonnabel2013stochastic}.
The different utility functions and order of play results in player-$i$ owning the eigenvector $\hat{v}_i$ associated with the $i^{th}$ largest eigenvalue $\lambda_i$.
In addition to the above formulation that assumes ordered play, \citet{gemp2020eigengame} empirically show that simultaneous play also converges to Nash equilibrium.

\section{NNMF as a multiplayer game}
Before we formulate NNMF as a game, we describe traditional algorithms used for it.

\subsection{Traditional algorithms for NNMF}

In it's most general form, NNMF is an optimization problem that involves minimizing some loss function $L(\mathbf{X}, \mathbf{WH})$ that penalizes the divergence between the original data matrix $\mathbf{X}$ and it's reconstruction $\mathbf{WH}$.

We will work with the following square Frobenius norm penalty for the rest of this manuscript, though more sophisticated objectives, including those that include various regularization terms, are possible \cite{berry2007algorithms}:
\begin{align}
\underset{w_{i,k}, h_{k,j} \geq 0} {\operatorname{minimize}}  || \mathbf{X} - \mathbf{W} \mathbf{H} ||^2_{F}
\end{align}
\label{nnmf-opt}

This optimization formulation is non-convex and obtaining globally optimum solutions is known to be NP-Hard \cite{vavasis2010complexity}.
However, interest in NNMF has led to the development of many empirically successful algorithms for finding local optima.
These algorithms fall into three general classes (see \citet{berry2007algorithms} for a thorough treatment): 

\paragraph{1. Non-negative Alternating Least squares (NALS): } 
While the optimization in equation \ref{nnmf-opt} is not convex in both $\mathbf{W}$ and $\mathbf{H}$, it is indeed convex in each term taken individually (\textit{biconvex}).
NALS solves a least-squares problem alternatively for $\mathbf{W}$ and $\mathbf{H}$, projecting the iterates into the positive orthant at each iteration, till convergence is observed. 

\paragraph{2. Projected Gradient (PG): }
Similar to NALS, PG performs a gradient ascent step on $\mathbf{W}$ and $\mathbf{H}$ alternatively, and projects the resulting iterates into the positive orthant at each iteration, till convergence is observed. 

\begin{algorithm}[tbhp]
\begin{algorithmic}[1]
    \STATE Given: data matrix, $\mathbf{X} \in \mathbb{R}^{I \times J}$, latent dimension $K$, maximum iterations $T$, randomly initialized $\mathbf{W}^{0} \in \mathbb{R}_{++}^{I \times K}$ and $\mathbf{H}^{0} \in \mathbb{R}_{++}^{K \times J}$.
    \FOR{$t = 1: T$}
        \STATE ${\displaystyle \mathbf {H} ^{t+1}\leftarrow \mathbf {H} ^{t} \odot {\frac {((\mathbf {W} ^{t})^{T}\mathbf {V} )}{((\mathbf {W} ^{t})^{T}\mathbf {W} ^{t}\mathbf {H} ^{t})}}}$
        \STATE ${\displaystyle \mathbf {W} ^{t+1}\leftarrow \mathbf {W} ^{t} \odot {\frac {(\mathbf {V} (\mathbf {H} ^{t+1})^{T})}{(\mathbf {W} ^{t}\mathbf {H} ^{t+1}(\mathbf {H} ^{t+1})^{T})}}}$
    \ENDFOR
    \STATE return $\mathbf{W}^t$ and $\mathbf{H}^t$ \\
     Note: $\odot$ and $\mathbf{X}/\mathbf{Y}$ are elementwise multiplication and division respectively.
\end{algorithmic}
\caption{Multiplicative updates \cite{lee1999learning}}
\label{algo:mw}
\end{algorithm}

\paragraph{3. Multiplicative updates (MU):}
In PG, the choice of gradient step size is typically done heuristically.
As an improvement, \citet{lee1999learning} popularized a variant of PG that uses an adaptive step size and results in a multiplicative update for $\mathbf{W}$ and $\mathbf{H}$ at each iteration.
The complete algorithm is given in Algorithm \ref{algo:mw}.

\subsection{Game-theoretic NNMF formulation}
\label{sec:nnmfgame}

\begin{figure}[tbhp]
\centering
\includegraphics[width=0.99\linewidth,interpolate=false]{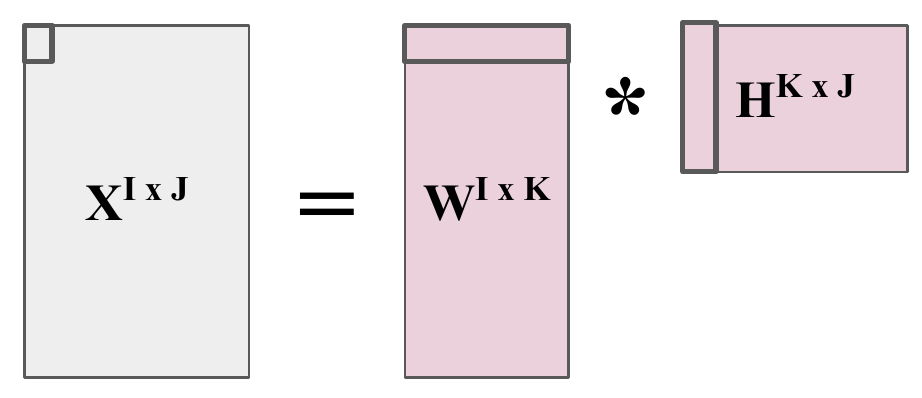}
\label{fig:nnmf}
\caption{\textbf{NNMF as a multiplayer game}: 
NNMF is formulated as a game between $I$ row-players, each owning one row of the $\mathbf{W}$ matrix, that play with $J$ column-players, each owning one column of the $\mathbf{H}$ matrix.
In each interaction, one row-player $i$ and one column-player $j$ maximize their respective utilities to better reconstruct one element of the data-matrix $\mathbf{X}_{[i, j]}$.
Additional games between the column players can be introduced to nudge the game towards desirable solutions, such as orthogonal rows for $\mathbf{H}$.
}
\end{figure}

To set up NNMF as a game, we define two sets of players, $I$ row-players and $J$ column-players and their respective utilities.
Each of the $I$ row-players owns a row $\mathbf{W}_{[i,:]}$ of the $\mathbf{W}$ matrix, and each the $J$ column-players owns a column $\mathbf{H}_{[:,j]}$ of the $\mathbf{H}$ matrix.
Therefore, each player owns a \mbox{length-$K$} non-negative vector.

Taking inspiration from the game formulation of PCA, we now define a utility function for each player, such that a player's local actions (i.e. best response) achieve a global optimization objective.
The primary interaction in this game is between row-players playing column-players and vice-versa.
The utility function in these interactions is given by: 
\begin{align}
u_{(i, j)} = -\mathbf{\ell}(i, j) = -\frac{1}{2} [\,\mathbf{X}_{[i,j]} - (\mathbf{W}_{[i,:]})^T \mathbf{H}_{[:,j]} ]\,^2     
\end{align}
\label{eq:nnmf}

In other words, the $i^{th}$ row-player and $j^{th}$ column-player \textit{cooperate} to try to reconstruct the $(i,j)^{th}$ element of the data matrix $\mathbf{X}$.
The NNMF game algorithm is summarized in Algorithm \ref{algo:nnmfgame}.

The formulation described up to this point can be regarded as a \textit{graphical game} on a bipartite graph with row-players and column players comprising the two disjoint sets.
We now relax the bipartite structure to a more general graphical game where there are interactions between some identical players, similar to replicator or winner-take-all dynamics from evolutionary game theory \cite{sandholm2010population}.  
Column players (owning columns of the $\mathbf{H}$ matrix) now have \textit{self-games} as a way to induce certain forms of regularization that aid interpretation of the resulting matrix $\mathbf{H}$.
For example, to encourage non-overlapping rows in $\mathbf{H}$, we could have a column player play against itself in such a way that the smallest element of the column-vector it owns is set to zero.
A more aggressive regularization strategy towards the same goal could be that all but the largest element of the column-vector stay non-zero after the self-interaction.
Henceforth, we call these strategies \texttt{J-min} and \texttt{J-max} respectively, 
In practice by instead of setting an element of the column-vector to zero, we multiply it by $0.99$ to nudge it in the right direction.


\begin{algorithm}[tbhp]
\begin{algorithmic}[1]
    \STATE Given: data matrix, $\mathbf{X} \in \mathbb{R}^{I \times J}$, latent dimension $K$, maximum iterations $T$, step size $\eta$, randomly initialized $\mathbf{W}^{0} \in \mathbb{R}_{++}^{I \times K}$ and $\mathbf{H}^{0} \in \mathbb{R}_{++}^{K \times J}$.
    \FOR{$t = 1: T$}
        \FOR{$i = 1: I$}
            \FOR{$j = 1: J$}
               \STATE $\nabla^{t}_{W} \leftarrow -[\,\mathbf{X}_{[i,j]} - (\mathbf{W}^{t}_{[i,:]})^T  \mathbf{H}^{t}_{[:,j]}]\, (\mathbf{H}^{t}_{[:,j]})^T$
               \STATE $\nabla^{t}_{H} \leftarrow -[\,\mathbf{X}_{[i,j]} - (\mathbf{W}^{t}_{[i,:]})^T  \mathbf{H}^{t}_{[:,j]}]\, (\mathbf{W}^{t}_{[i,:]})^T$
               \STATE $\mathbf{W}^{t+1}_{[i,:]} \leftarrow max [\,\mathbf{0}, \mathbf{W}^{t}_{[i,:]} - \eta \nabla^{t}_{W} ]\, $
               \STATE $\mathbf{H}^{t+1}_{[:,j]} \leftarrow max [\,\mathbf{0}, \mathbf{H}^{t}_{[:,j]} - \eta  \nabla^{t}_{H} ]\, $
            \ENDFOR
        \ENDFOR
    \ENDFOR
    \STATE return $\mathbf{W}^{t}$ and $\mathbf{H}^{t}$ \\
\end{algorithmic}
\caption{NNMF game (simultaneous play)}
\label{algo:nnmfgame}
\end{algorithm}

\subsection{Parallelization and scaling properties}
The game-theoretic formulation of NNMF has several favorable parallelization and scaling properties:

\paragraph{Data sharding:} 
Each row-player and each column-player only require access to their respective row or column of the data matrix $\mathbf{X}$.
In Algorithm \ref{algo:nnmfgame} only a single element of the data matrix is required for every row-column player interaction.
If each row-player and each column-player make a local copy of their associated row or column of the data-matrix, that would be sufficient for all possible row-column interactions they have.
Furthermore, if we assume ordered play in row-column interactions, only the leader need hold the data matrix.
They can then share the required ground-truth matrix element with the other player during the interaction.  

In our formulation, we assume that each (row or column) player is on a separate computational core.
However, the algorithm is indifferent to the actual distribution of players across computational cores. 
This distribution can be arbitrary or be dictated by some optimal trade-off between inter-core communication cost and per-core memory cost. 

\paragraph{Computational cost:} 
The computational cost, in floating point operations (FLOPS), of multiplying one $I \times K$ matrix with a $K \times J$ matrix is $O(I J K)$.

Therefore one MU iteration (Algorithm \ref{algo:mw}) costs \mbox{ $O(IJK)$ [${((\mathbf {W} ^{t})^{T}\mathbf {V} )}$]} + $O(IK^2 + JK^2)$ [${((\mathbf {W} ^{t})^{T}\mathbf {W} ^{t}\mathbf {H} ^{t})}$] + $O(KJ)$ [$\mathbf{X/Y}$] + $O(KJ)$ [$\odot$] for the $\mathbf {H}$ update. 
A similar calculation applies to the $\mathbf {W}$ update.

One NNMF iteration (Algorithm \ref{algo:nnmfgame}) involves $I J$ row-column player interactions, each with a cost \mbox{$O(K^2)$ [$\nabla^{t}_{W} \leftarrow -[\,\mathbf{X}_{[i,j]} - (\mathbf{W}^{t}_{[i,:]})^T  \mathbf{H}^{t}_{[:,j]}]\, (\mathbf{H}^{t}_{[:,j]})^T$]} + \mbox{$O(K)$ [ $\mathbf{W}^{t+1}_{[i,:]} \leftarrow \mathbf{W}^{t}_{[i,:]} - \eta \nabla^{t}_{W} $ ]} for the $W_{[i,:]}$ update. 
A similar calculation applies for the $\mathbf {H}$ update.
This is comparable with the cost of the MU update for small $K$, while being highly parallelizable.

\section{Empirical analysis}
To validate our new game-theoretic formulation of NNMF, we empirically compare it with the traditional Multiplicative Updates (MU) algorithm using synthetic datasets. 

\begin{figure}[tbhp]
\centering
\includegraphics[width=0.99\linewidth,interpolate=false]{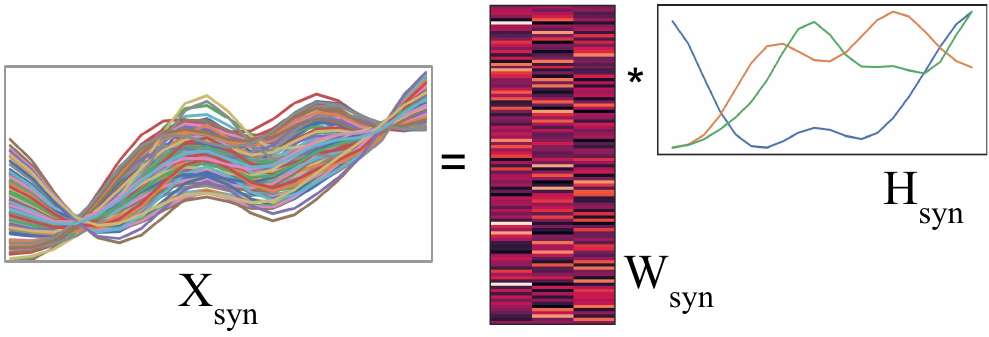}
\caption{Synthetic data for experiments: Data matrix $\mathbf{X_{syn}} \in \mathbb{R}^{I \times J}$ reconstructed from matrix multiplication of $\mathbf{H_{syn}} \in \mathbb{R}_{++}^{K \times J}$ and $\mathbf{W_{syn}} \in \mathbb{R}_{++}^{I \times K}$. }   
\label{fig:syn}
\end{figure}

\paragraph{Synthetic dataset generator:} 
For all experiments, we set the number and dimension of our observations to be $I = 100$ and $J = 20$ respectively, resulting in a data matrix $\mathbf{X_{syn}} \in \mathbb{R}^{I \times J}$.
We set the true number of latent factors to $K = 3$, and generate a random Uniform$(0,1)$ non-negative mixing matrix $\mathbf{W_{syn}} \in \mathbb{R}_{++}^{I \times K}$.
We generate the non-negative basis matrix $\mathbf{H_{syn}} \in \mathbb{R}_{++}^{K \times J}$ by applying a Gaussian filter (window-length=3) to the rows of a ${K \times J}$ random matrix with elements drawn from a Uniform$(0,1)$ distribution.
$\mathbf{X_{syn}}$ is then obtained multiplying $\mathbf{W_{syn}}$ and $\mathbf{H_{syn}}$ (Figure \ref{fig:syn}).

\paragraph{Experiment and Results:}
\begin{figure}[tbhp]
\centering
\includegraphics[width=0.99\linewidth,interpolate=false]{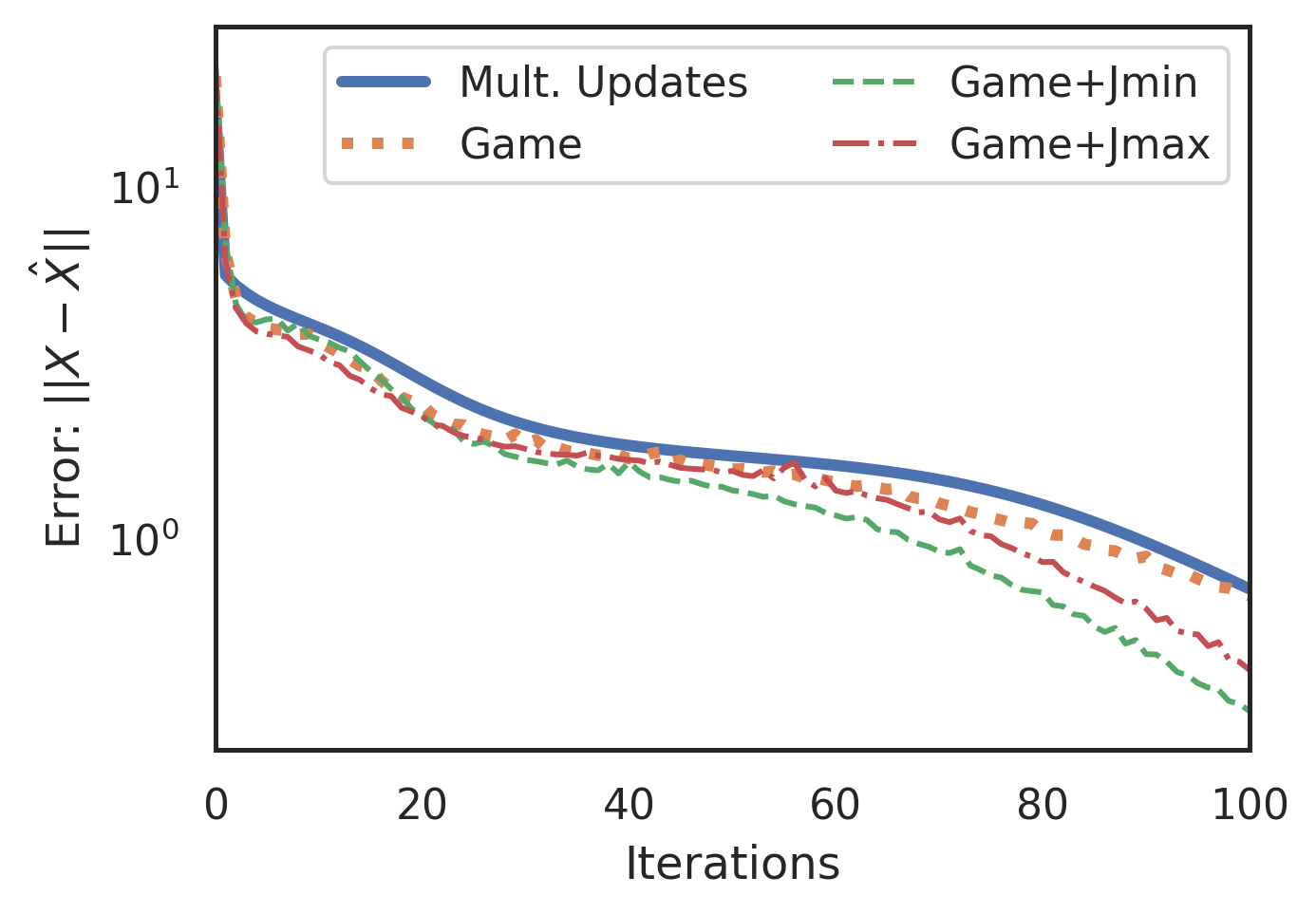}
\includegraphics[width=0.80\linewidth,interpolate=false]{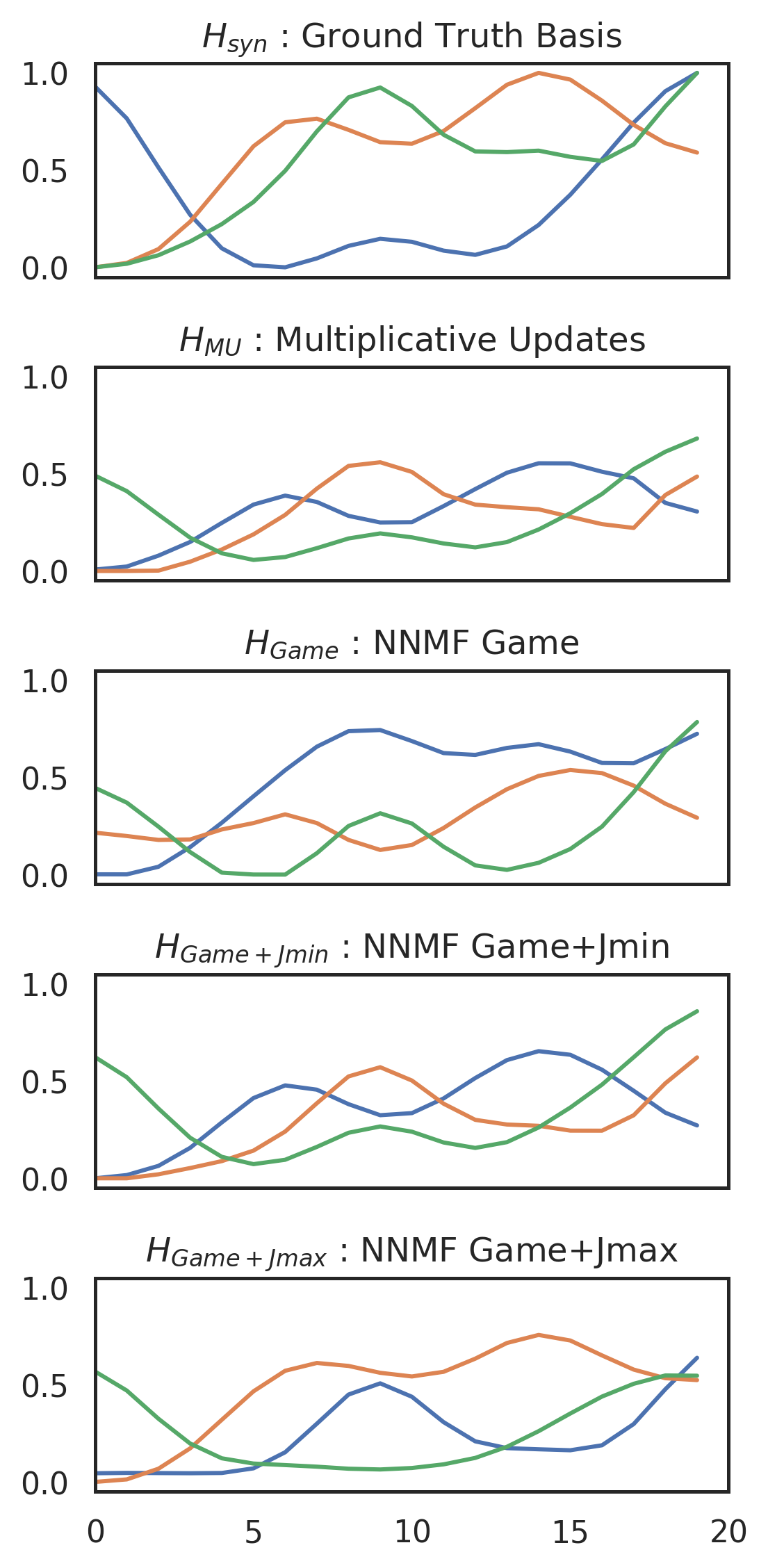}
\caption{\textbf{Convergence of algorithm variants [Top]}: 
Game-theoretic variants of NNMF ($\eta = 0.001$) have convergence rates comparable to the traditional Multiplicative Updates (MU) algorithm. \\
\textbf{Ground truth vs. Recovered Bases [Lower]}: 
Basis matrices $\mathbf{H_{syn}}$ recovered from MU and Game variants look similar, up to permutations of order (colors). 
The \texttt{J-min} and \texttt{J-max} variants only show slight effects of regularization, possibly because the ground-truth basis is not exactly orthogonal.
}    
\label{fig:convergence_recoveredH}
\end{figure}

\begin{figure*}[tbhp]
\centering
\includegraphics[width=0.99\linewidth,interpolate=false]{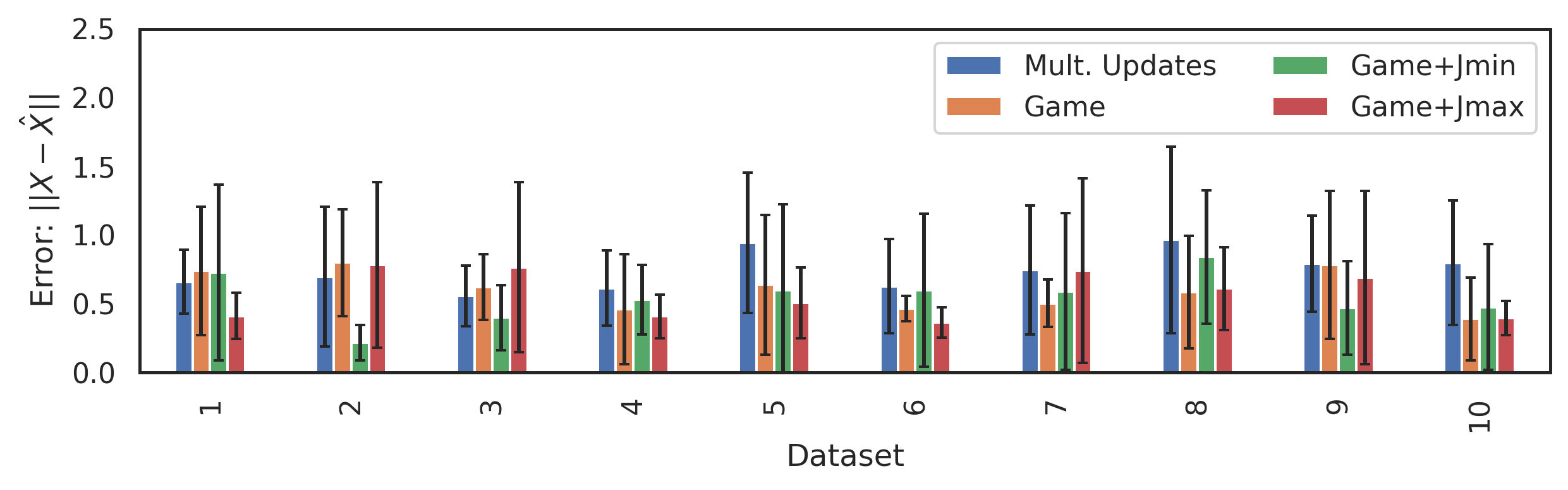}
\caption{\textbf{Reconstruction error of algorithm variants evaluated over multiple randomly generated synthetic datasets}: 
Game-theoretic NNMF variants result in reconstruction errors that are not consistently significantly different compared to reconstruction errors from the traditional MU algorithm. 
Error bars represent \mbox{$\pm$ 1 s. d.}, calculated over 4 random initializations.}   
\label{fig:rounds}
\end{figure*}

\begin{figure}
\centering
\includegraphics[width=0.99\linewidth,interpolate=false]{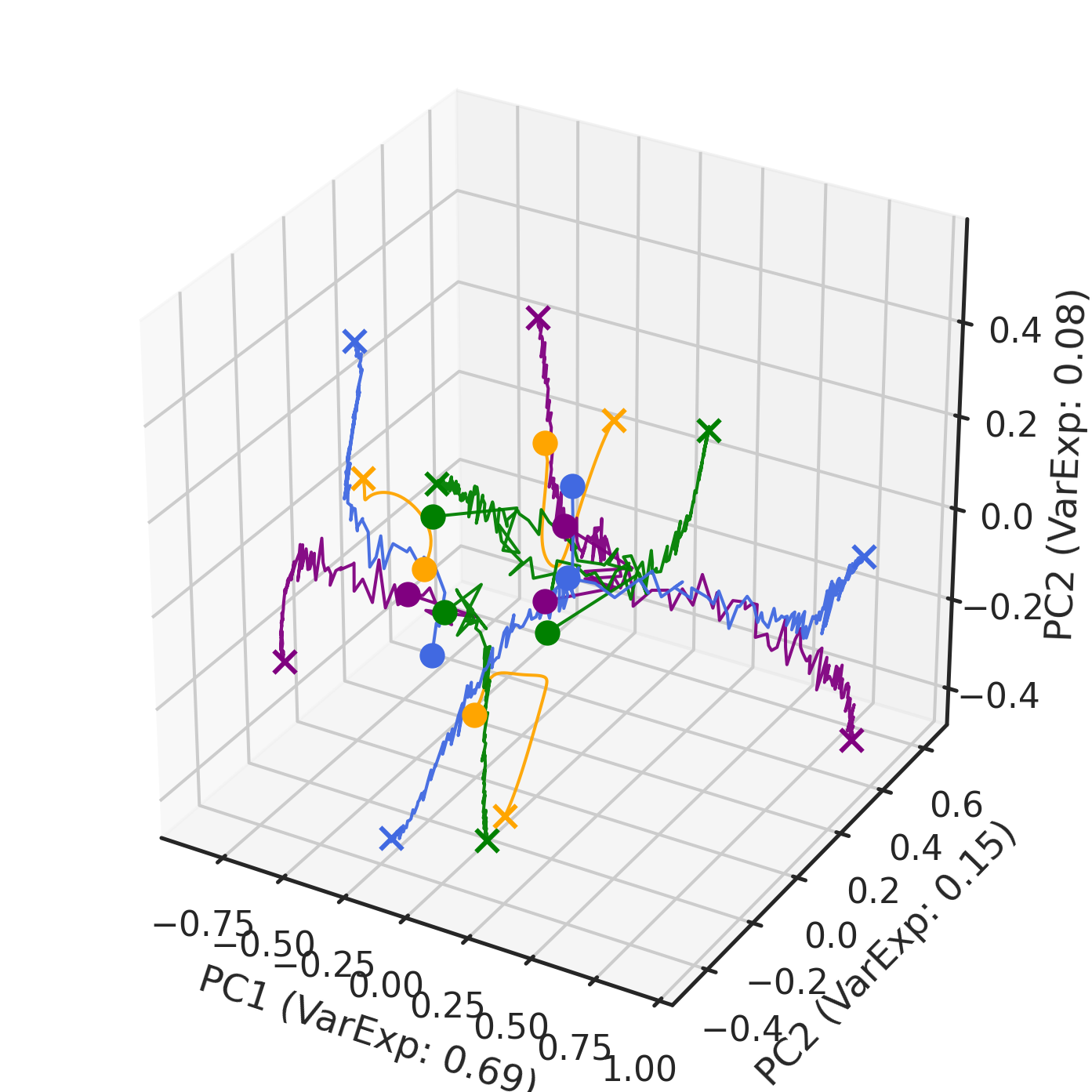}
\caption{\textbf{Low dimensional projection of iterate trajectories for $\mathbf{\hat {H}_{syn}}$ basis matrix rows:}
Trajectories for Multiplicative Updates (MU) [orange], NNMF game vanilla [purple], \texttt{Jmin} [green], and \texttt{Jmax} [blue] variants (using same initialization as in \mbox{Figure \ref{fig:convergence_recoveredH}}).
Each algorithm has $K=3$ trajectories associated with it, one for each row of the recovered basis matrix $\mathbf{\hat {H}_{syn}}$, each starting at a $\bullet$ marker and ending at a $\times$ marker.
Trajectories are observed to be relatively smooth for the (adaptive step-size) MU algorithm and relatively jagged for the (fixed step-size) NNMF game variants.
Though iterates seem to be moving away from each other, it is not clear if this can be regarded as a signature of congestion as claimed for iterates in the PCA game \citet{gemp2020eigengame}.
Low dimensional space is generated from applying PCA on all $\mathbf{\hat {H}_{syn}}$ basis vectors from all algorithms stacked together.
PC1, PC2 and PC3 explain 69, 15 and 8 percent of the total variance of the data respectively.
} 
\label{fig:traces}
\end{figure}

We generate a synthetic dataset and compare its decomposition using MU algorithm to the one obtained using our game-theoretic NNMF algorithm and it's two regularized variants, \texttt{J-min} and \texttt{J-max}, Section \ref{sec:nnmfgame} (using a fixed gradient step size of $\eta = 0.001$ across all NNMF executions).  
We find that the convergence trajectories of the MU and NNMF-Game algorithms are comparable, and that the recovered basis matrices $\mathbf{\hat {H}}_{syn}$ look quite similar (Figure \ref{fig:convergence_recoveredH}). 

Next, we examine the evolution of the non-negative basis (row) vectors that comprise the basis matrix $\mathbf{\hat {H}}_{syn}$ by plotting a low-dimensional projection of the trajectories of the iterates (Figure \ref{fig:traces}).
This reveals trajectories that appear smooth for the (adaptive step size) MU algorithm and relatively jagged for our (fixed step size) NNMF game variants.

Finally, we compare the reconstruction performance ($ Error = || \mathbf{X} - \mathbf{\hat{X}} ||_{2} $) of these algorithms across 10 new independently randomly generated synthetic datasets, with 4 randomly initialized episodes for each. 
We find that the performance of the game-theoretic NNMF variants and that of the traditional MU algorithm is not significantly different (Figure \ref{fig:rounds}).

\section{Discussion}
We present a game formulation of the popular Non-negative Matrix Factorization algorithm with favorable parallelization and scaling properties.
The game structure admits variants that can induce certain forms of interpretability-enhancing regularization such as encouraging rows in the recovered basis matrix $\mathbf{\hat {H}}$ to be orthogonal. 
We provide empirical evidence that our algorithms have convergence and reconstruction performance similar to those of the traditional Multiplicative Updates (MU) \cite{lee1999learning} algorithm.
Visual inspection of convergence trajectories for rows of the recovered basis matrix $\mathbf{\hat {H}}$ reveals that iterates tend to vary relatively smoothly for the (adaptive step-size) MU algorithm but relatively jaggedly for our (fixed step-size) NNMF game variants.

For future work, we plan to formally analyse the convergence properties of the graphical game variants introduced in this paper, and explore adaptive learning rate and additional regularization strategies that could result in smoother convergence trajectories and additional desirable properties in the recovered factor matrices.
Exploring extensions to the more general Tensor Factorization setting, and connections to Message Passing inference algorithms are other directions that seem promising.  

\newpage
\section*{Acknowledgements}
We thank Prof. Lillian Ratliff for helpful discussions.

\bibliography{report}

\begin{thebibliography}{10}
\providecommand{\natexlab}[1]{#1}
\providecommand{\url}[1]{\texttt{#1}}
\expandafter\ifx\csname urlstyle\endcsname\relax
  \providecommand{\doi}[1]{doi: #1}\else
  \providecommand{\doi}{doi: \begingroup \urlstyle{rm}\Url}\fi

\bibitem[Berry et~al.(2007)Berry, Browne, Langville, Pauca, and
  Plemmons]{berry2007algorithms}
Berry, M.~W., Browne, M., Langville, A.~N., Pauca, V.~P., and Plemmons, R.~J.
\newblock Algorithms and applications for approximate nonnegative matrix
  factorization.
\newblock \emph{Computational statistics \& data analysis}, 52\penalty0
  (1):\penalty0 155--173, 2007.

\bibitem[Bonnabel(2013)]{bonnabel2013stochastic}
Bonnabel, S.
\newblock Stochastic gradient descent on {R}iemannian manifolds.
\newblock \emph{IEEE Transactions on Automatic Control}, 58\penalty0
  (9):\penalty0 2217--2229, 2013.

\bibitem[Gemp et~al.(2020)Gemp, McWilliams, Vernade, and
  Graepel]{gemp2020eigengame}
Gemp, I., McWilliams, B., Vernade, C., and Graepel, T.
\newblock Eigengame: {PCA} as a {N}ash {E}quilibrium.
\newblock \emph{arXiv preprint arXiv:2010.00554}, 2020.

\bibitem[Gemp et~al.(2021)Gemp, McWilliams, Vernade, and
  Graepel]{gemp2021eigengame}
Gemp, I., McWilliams, B., Vernade, C., and Graepel, T.
\newblock Eigengame unloaded: When playing games is better than optimizing.
\newblock \emph{arXiv preprint arXiv:2102.04152}, 2021.

\bibitem[Jolliffe(2002)]{jolliffe2002principal}
Jolliffe, I.~T.
\newblock Principal components in regression analysis.
\newblock In \emph{Principal Component Analysis}. Springer, 2002.

\bibitem[Lee \& Seung(1999)Lee and Seung]{lee1999learning}
Lee, D.~D. and Seung, H.~S.
\newblock Learning the parts of objects by non-negative matrix factorization.
\newblock \emph{Nature}, 401\penalty0 (6755):\penalty0 788--791, 1999.

\bibitem[Pauca et~al.(2004)Pauca, Shahnaz, Berry, and Plemmons]{pauca2004text}
Pauca, V.~P., Shahnaz, F., Berry, M.~W., and Plemmons, R.~J.
\newblock Text mining using non-negative matrix factorizations.
\newblock In \emph{Proceedings of the 2004 SIAM International Conference on
  Data Mining}, pp.\  452--456. SIAM, 2004.

\bibitem[Sandholm(2010)]{sandholm2010population}
Sandholm, W.~H.
\newblock \emph{Population games and evolutionary dynamics}.
\newblock MIT press, 2010.

\bibitem[Stein-O’Brien et~al.(2018)Stein-O’Brien, Arora, Culhane, Favorov,
  Garmire, Greene, Goff, Li, Ngom, Ochs, et~al.]{stein2018enter}
Stein-O’Brien, G.~L., Arora, R., Culhane, A.~C., Favorov, A.~V., Garmire,
  L.~X., Greene, C.~S., Goff, L.~A., Li, Y., Ngom, A., Ochs, M.~F., et~al.
\newblock Enter the matrix: factorization uncovers knowledge from omics.
\newblock \emph{Trends in Genetics}, 34\penalty0 (10):\penalty0 790--805, 2018.

\bibitem[Vavasis(2010)]{vavasis2010complexity}
Vavasis, S.~A.
\newblock On the complexity of nonnegative matrix factorization.
\newblock \emph{SIAM Journal on Optimization}, 20\penalty0 (3):\penalty0
  1364--1377, 2010.

\end{thebibliography}
\bibliographystyle{icml2021}
\end{document}